\documentclass[epj]{webofc}

\pdfoutput=1

\usepackage[varg]{txfonts}   
\wocname{EPJ Web of Conferences}
\woctitle{ICNFP 2016}
\usepackage{color,graphicx}   
\definecolor{red}{rgb}{0.8,0,0}
\definecolor{violet}{rgb}{0.4,0,0.4}
\definecolor{green}{rgb}{0,0.5,0.0}
\definecolor{navy}{rgb}{0.0,0.0,0.6}
\definecolor{orange}{rgb}{0.8,0.2,0.0}
\newcommand{\bea}{\begin{eqnarray}}
\newcommand{\eea}{\end{eqnarray}}

\def\apj{ApJ~}%
\def\apjl{ApJ Lett.~}%
\def\aap{A\&A~ }%
\def\jcap{J. Cosmo. Astropart. Phys.~}%
\def\mnras{MNRAS~}%
\def\prc{Phys.~Rev.~C~}%
\def\prd{Phys.~Rev.~D~}%
\def\nat{Nature~}%
\def\physrep{Phys.~Rep.~}%
\begin{document}
\selectlanguage{english}
\title{Exploring phases of dense  QCD with compact stars}
\author{Armen Sedrakian\inst{1}\fnsep\thanks{\email{sedrakian@th.physik.uni-frankfurt.de}} 
}
\institute{
  Institute for Theoretical  Physics, J.~W.~Goethe-University, D-60438 Frankfurt-Main, Germany 
}

\abstract{%
  I review a number of recent developments in the physics of compact
  stars containing deconfined quark matter, including (a)~their
  cooling with possible phase transition from a fully gapped to a
  gapless phase of QCD at low temperatures and large isospin; (b)~the
  transport coefficients of the 2SC phase and the role played by the
  Aharonov-Bohm interactions between flux-tubes and unpaired fermions;
  (c)~rapidly rotating compact stars and spin-down and spin-up
  induced phase transition between hadronic and QCD matter as well as
  between different phases of QCD.}
\maketitle
\section{Introduction}
\label{sec:intro}
A unique way to discern the phase structure of dense quantum
chromodynamics (QCD) is offered by the phenomenology of compact
stars. Their global parameters, such as the mass, the radius, the
moment of inertia etc. depend sensitively on the equation of state
(i.e. pressure vs density relation at zero temperature; the many-body
methods to derive an equation of state appropriate for compact stars
are reviewed, for example, in Refs.~\cite{WeberBook,Sed07}; for recent developments see Ref.~\cite{LP16}). The
recent measurements of pulsars masses in binary orbits with white
dwarfs inferred masses in the range 1.8 - 2.0 solar masses (hereafter
$M_{\odot}$)~\cite{2010Natur.467.1081D,2013Sci...340..448A,2016arXiv161103658B}.
These observations suggest that the equation of state of densest
regions of compact stars must be moderately stiff. There are
observational indications that the pulsar radii are not very large,
possibly in the range of $11\le R\le 14$ km, which suggest that at
lower densities the equation of state cannot be too stiff. The mass
bounds quoted above were important in constraining various models of
equations of state of dense matter in recent years.

The information gained from the global parameters of compact
stars is however limited, as the equation of state may be insensitive to the
detailed structure of QCD phases, which is dictated by the form of the
quasiparticle spectrum in the vicinity of the Fermi surface. An
obvious example is the pairing, which introduces a gap in the spectrum
of quasiparticles and leads to the phenomenon of superconductivity of
quarks~\cite{1984PhR...107..325B,2000hep.ph...11333R,2001ARNPS..51..131A,2008RvMP...80.1455A}. The cooling and transport in
compact stars is sensitive to such details and most of this review is
dedicated to the understanding of these properties of various phases
of dense QCD.

This review is structured as follows. Section~\ref{sec:cooling}
discusses the cooling of compact stars with quark cores, in particular
the differences in the neutrino emission from phases with and without
fully gapped Fermi spheres. Section~\ref{sec:transport} reviews the
recent computations of the transport coefficients of two-flavor cold
quark matter where two of the colors are paired (the so-called 2SC
phase). The possible role of color-magnetic flux tubes is also
discussed. Finally, in Sec.~\ref{sec:spin} we discuss the possibility
of observing a phase transition between the hadronic and
superconducting quark matter as well as phase transitions between two
different phases of dense QCD in spinning up and down compact
stars. Some final remarks are given in Sec.~\ref{sec:conclusions}.  An
earlier account of the topics listed above can be found in
Ref.~\cite{2013arXiv1301.2675S}. This review does not provide a
comprehensive coverage of the subjects listed above and is
focused on the research of the author.

\section{Cooling of compact stars}
\label{sec:cooling}

Currently the deconfinement of baryons into quark is not well
understood, consequently the densities at which such a phase
transition can take in compact stars is uncertain. The scale at which
we may expect such a transition is set by the characteristic size of a
baryon, which should be $\le 0.84$ fm - the measured charge radius of
the proton. Once the interparticle distance becomes of the order of
this scale the baryons will lose their identity, as their
wave-functions will start to overlap. Furthermore, the phase diagram
of quark matter at not very high densities is not precisely known,
because at the densities relevant to compact stars the perturbative QCD
is not valid and non-perturbative methods (such as lattice QCD)
face technical obstacles.

Cooling of compact stars can provide indirect information on the
properties of quark phases if the density of deconfinement is indeed
reached in some heavier stars. Such stars are known as {\it hybrid
  compact star}.  In this Section we summarize the results of the studies of
cooling of hybrid compact stars carried out in
Refs.~\cite{2011PhRvD..84f3015H,2013A&A...555L..10S,2016EPJA...52...44S}.
As part of this program, the rapid cooling of the compact star in
Cassiopea A (hereafter Cas A) was modelled as a phase transition
within the QCD phase diagram~\cite{2013A&A...555L..10S}.  The latest
updated data~\cite{2013ApJ...777...22E} that covers the 10 year period
from 2003 to 2013 was fitted in Ref.~\cite{2016EPJA...52...44S}.  The
unprecedented fast cooling of the compact star in Cas A, in fact,
requires fast transient cooling model for this object consistent with
the estimate of Ref.~\cite{2013ApJ...777...22E}, which indicates a
decline in the temperature of the star $2.9\pm 0.9\%$ over 10 years of
observation.  However, this data cannot be interpreted unambiguously
for a number of reasons, see the discussion in
Refs.~\cite{2013ApJ...777...22E,2013ApJ...779..186P}.

Before turning to the QCD based description of cooling of compact
stars and the Cas A case, it should be mentioned that a number of
alternative models describe theoretically the Cas A cooling behaviour
on the basis of physics unrelated to QCD degrees of freedom.
Nucleonic stellar models (i.e., those containing neutrons, protons,
and electrons) attribute the rapid cooling to the onset of Cooper
pair-breaking process in the neutron superfluid
component~\cite{2011PhRvL.106h1101P,2011MNRAS.412L.108S,2015MNRAS.446.3621S,2015PhRvC..91a5806H,2013ApJ...779L...4N,2014arXiv1411.6833L}.
An alternative nucleonic model of Ref.~\cite{2012PhRvC..85b2802B} uses
rates of modified Urca and bremsstrahlung processes that are enhanced
compared to rates used in the minimalistic models quoted above by
several orders of magnitude due to a softening of pionic modes, as
initially discussed in Ref.~\cite{1997A&A...321..591S}.  A further
model requires a fine-tuned onset of Urca process in rotating compact
stars~\cite{Negreiros}.

We consider models of hybrid compact stars which contain a color
superconducting quark core surrounded by a nucleonic envelope.  These
phases are separated by a sharp interface, see
Ref.~\cite{2012A&A...539A..16B}.  Stars with such hybrid structure
naturally correspond to the most massive members of the sequence of
stellar equilibria modelled with an equation of state which contains a
phase transition from nucleonic to quark matter. These massive members
must be heavy enough to account for the  inferred
masses of
pulsars~\cite{2010Natur.467.1081D,2013Sci...340..448A,2016arXiv161103658B}.

Non-superconducting quarks cool matter via the direct Urca processes
$d\to u + e+ \bar \nu$ and $ u + e\to d+ \nu$ at a rate which is much
larger than for the ordinary nucleonic matter with low proton
fraction~\cite{2011PhRvD..84f3015H,2000PhRvL..85.2048P,2005PhRvD..71k4011A,2006PhRvD..74g4005A,2005PhRvC..71d5801G}.
However, cold quark matter is a (color)
superconductor~\cite{1984PhR...107..325B,2000hep.ph...11333R,2001ARNPS..51..131A,2008RvMP...80.1455A}.
Because of $\beta$-equilibrium and non-zero strange quark mass, the
Fermi surfaces of up and down quarks are shifted by amount which is of
the order of the electron chemical potential or strange quark
mass. This implies that new superconducting phases will emerge, in which
the Cooper pairing differs from the ordinary Bardeen-Cooper-Schrieffer
pairing.  Examples are the gapless two-flavor
phases~\cite{2003PhLB..564..205S,2003PhRvD..67h5024M} or the the
crystalline color-super\-conducting phase
\cite{2001PhRvD..63g4016A,2014RvMP...86..509A}.  The latter phase has
a number of realizations, because there are many ways to break the
spatial symmetries by allowing Cooper pairs to carrying finite
momentum.  Ref.~\cite{2016EPJA...52...44S} assumed the so-called
Fulde-Ferrell phase (hereafter FF phase), which postulates a single
plane wave modulation of the condensate. As far as the cooling is
concerned , the general properties of crystalline phases - the
existence of gapless
excitations~\cite{2001PhRvD..63g4016A,2014RvMP...86..509A,2009PhRvD..80g4022S,2010PhRvD..82d5029H}
- is present already in the FF model.

\begin{figure}[t]
\centering 
\sidecaption 
\includegraphics[width=7cm]{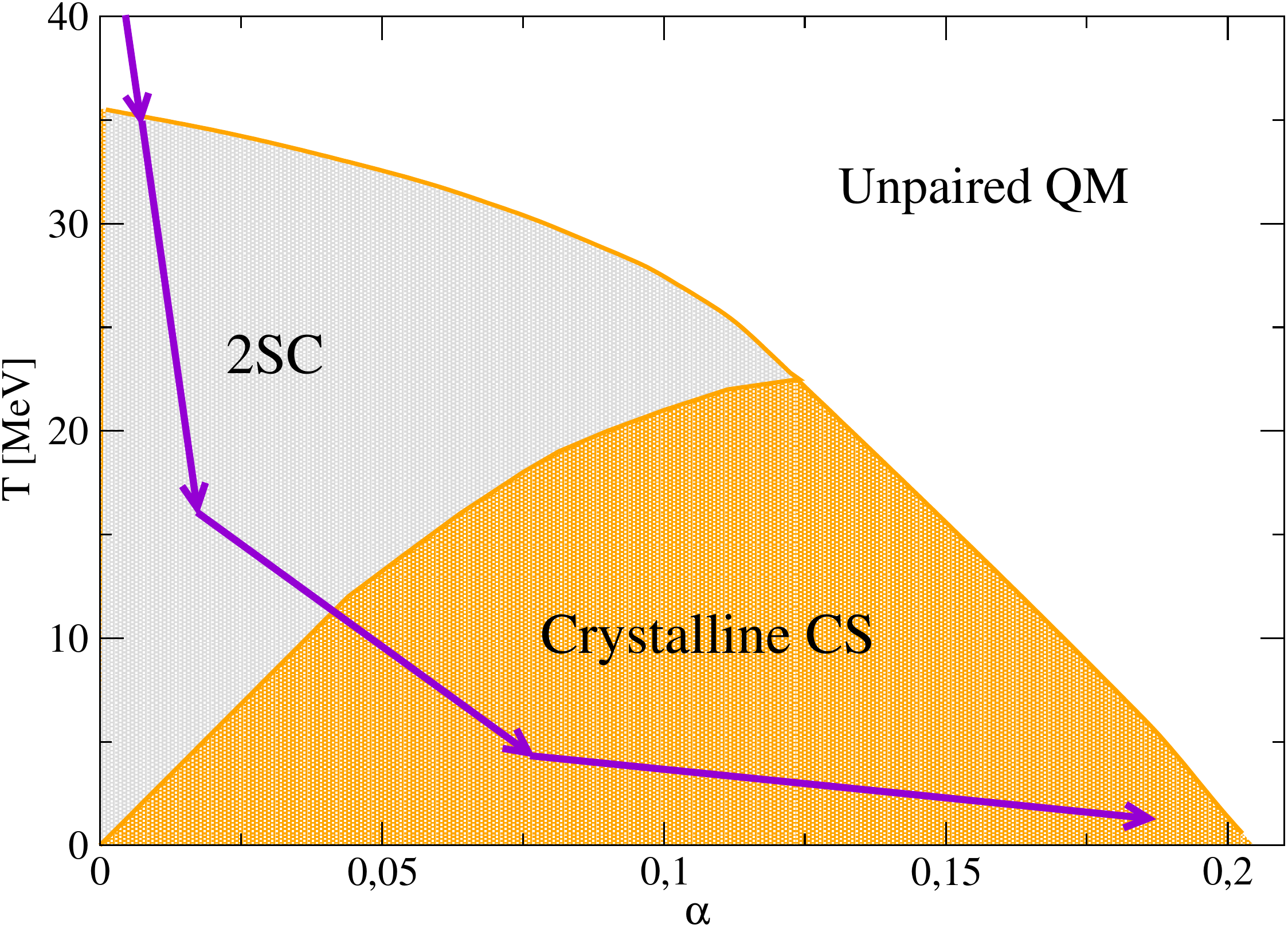}
\caption{The 2SC and FF phases in the phase diagram of two-flavor
  quark matter spanned by the isospin asymmetry $\alpha =
  (n_d-n_u)/(n_u+n_d)$, where $n_d$ and $n_u$ are the number densities
  of $d$ and $u$ flavors of quarks, and temperature.  The path that a
  compact star traverses during the cooling process is shown by
  arrows.}
\label{fig-1}       
\end{figure}
A compact star born in a supernova explosion is hot and matter is
nearly isospin symmetrical. In the subsequent minutes to hours the
matter cools rapidly and the isospin asymmetry increases to the values
characteristic for neutron stars. For temperatures $T\le T_c$, where
$T_c$ is the critical temperature of phase transition, and not too
high isospin asymmetries one finds the homogeneous 2SC phase in which
the rotational/translational symmetries are unbroken,
see~Fig.~\ref{fig-1}. For low temperature and large asymmetries one
finds that the ground state corresponds to the FF state.

The consequences of the transition from the symmetrical two-flavor BCS
phase to the crystalline phase at some temperature $T^*$ for the
cooling of hybrid stars with application to the Cas A case were studied in
Ref.~\cite{2016EPJA...52...44S}.  Fits were performed to obtain the
value of this temperature. Because the cross-flavor pairing occurs
between the green and red colored quarks and the blue quarks do not
participate in this type of the pairing a second parameter - the gap
for blue-colored quarks $\Delta_b$ - was introduced in the color
${\bf 6}_S$ and flavor ${\bf 3}_S$ channel. Note that this
same-flavor and same-color pairing is unaffected by the
flavor asymmetry~\cite{2003PhRvD..67e4018A}. An addition parameter $w$
is introduced to account for the finite time-scale of the phase
transition, see Ref.~\cite{2016EPJA...52...44S} for details.
\begin{figure}[bth]
\centering 
\begin{minipage}{6.0cm}
\includegraphics[width=6cm]{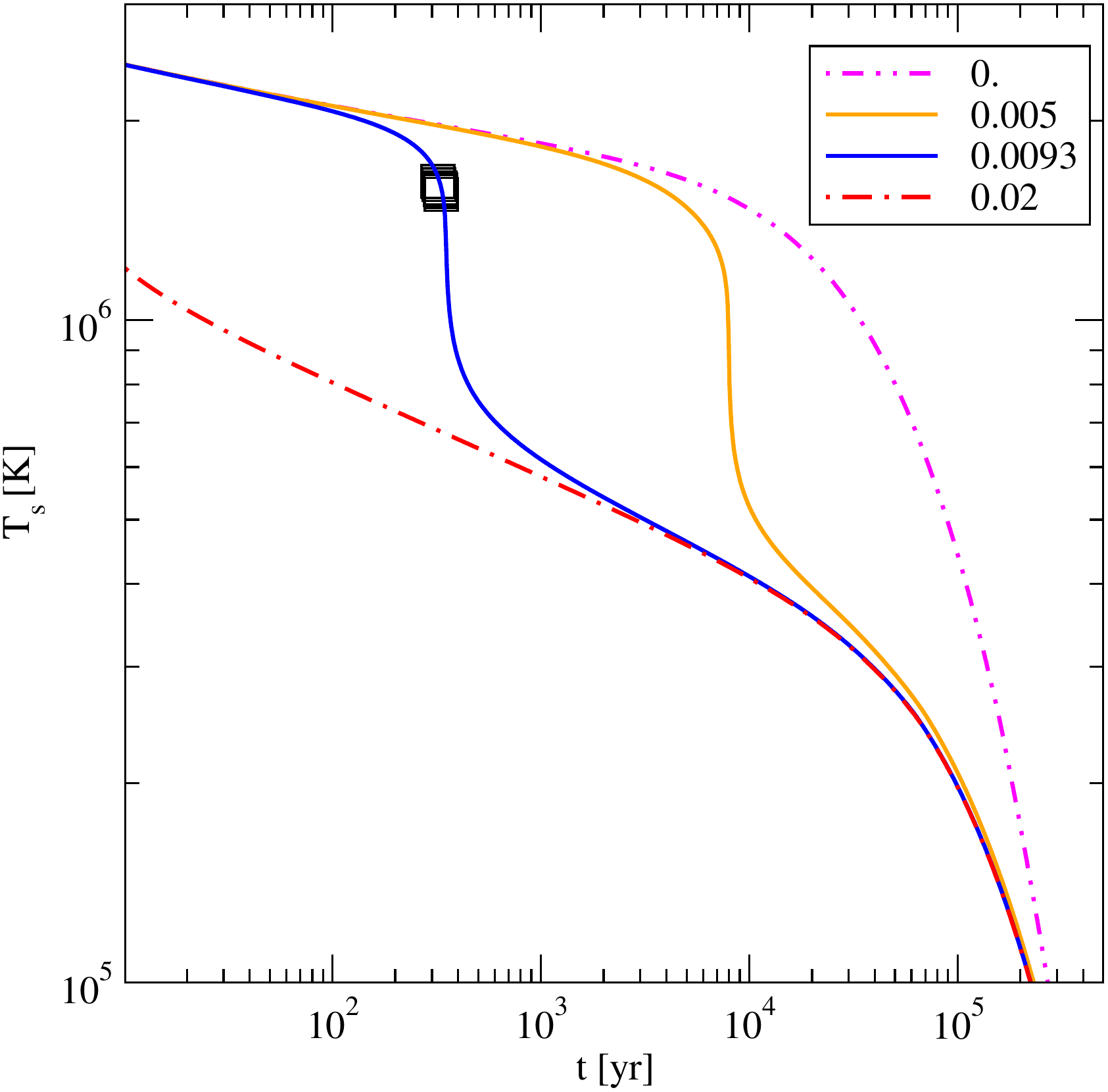}
\end{minipage}
\begin{minipage}{6.0cm}
\vskip -0.2cm 
\includegraphics[width=6cm]{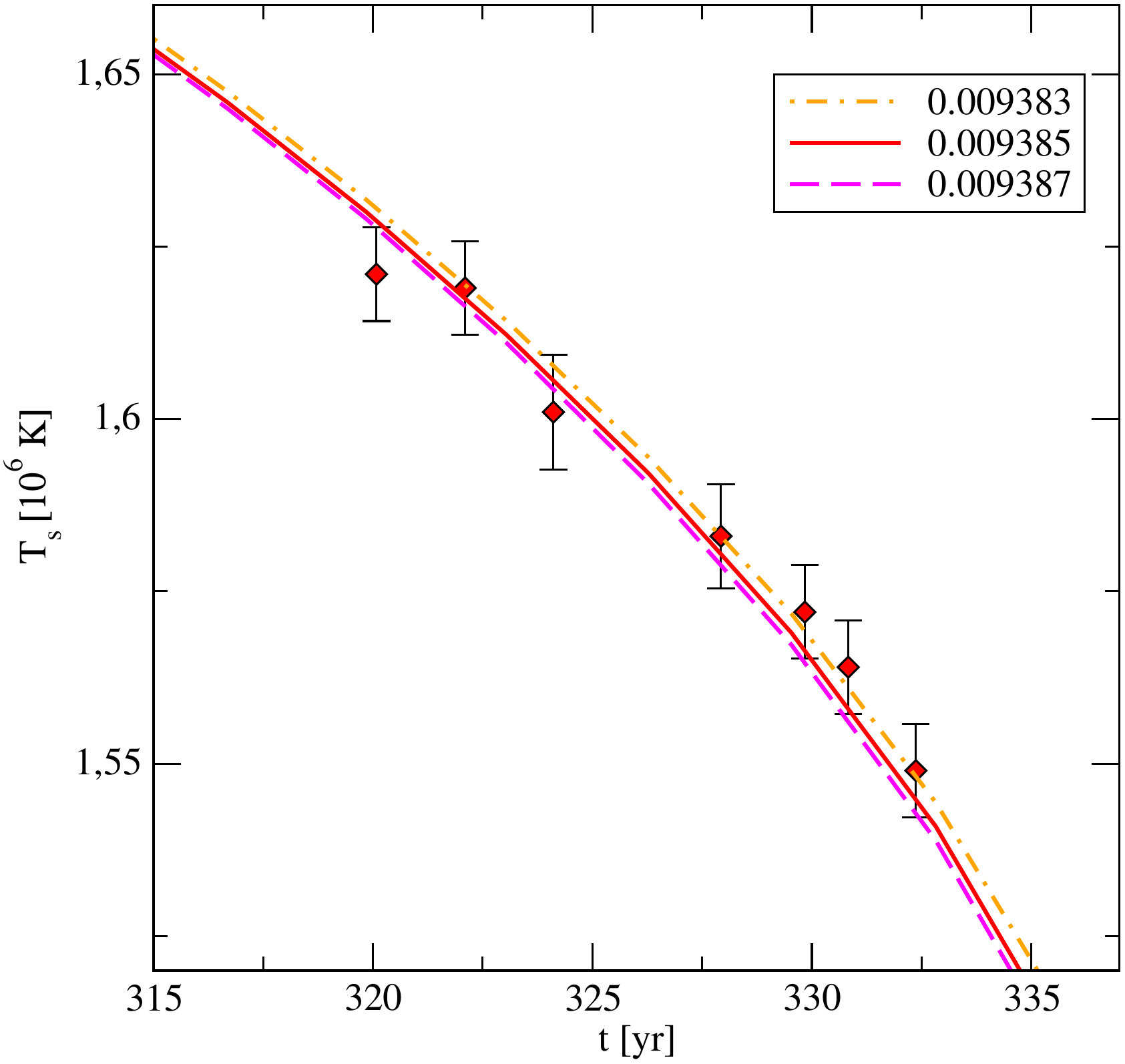}
\end{minipage}
\caption{Dependence of the surface temperature of a CSs given in K on 
  time in year. The lines correspond to different phase transition 
  temperature $T^*$ for fixed values of the width $w = 0.2$ and the 
  blue quark pairing gap $\Delta_b= 0.15$ (both in MeV). The labels 
  correspond to values of $T^*$ in MeV.  The Cas A data is shown by 
  squares.}
\label{fig-2}       
\end{figure}
Illustrative results are shown in Fig.~\ref{fig-2}, where we
display the dependence of the (redshifted) surface temperature on time
for a stellar model with mass $M/M_{\odot} = 1.93$. The parameter $T^*$ is
varied as indicated in the figure whereas the parameters $w$ and
$\Delta_b$ are held fixed.

For small $T^*$ the quark core has no influence on the cooling because
the neutrino emission from the red-green and blue condensates is
suppressed by the pairing gaps in these channels. In the opposite case
of high values of $T^*\sim 0.2$ MeV the transition to the FF phase
occurs early in time and fast neutrino emission via the direct Urca
process cools the star rapidly below the value measured for Cas
A. After fine tuning of the phase-transition temperature to the value
$T^*=0.009385 $ MeV the temperature of Cas A can be reproduced
accurately (Fig.~\ref{fig-2} left panel). As seen from the right panel
of Fig.~\ref{fig-2} a further tuning of the parameter $T^*$ allows us
to fit the details of the data with good accuracy.

Thus, we conclude that the rapid cooling of the compact star in Cas A
can be accounted for via the phase transition from the 2SC to FF-type
phase. It is interesting that the phase transition takes place {\it
  within the phase diagram of QCD} and is the consequence of the
ordering of various superconducting phases in the temperature,
density, and isospin spaces. Such ordering was observed in numerous
studies of other {\it population imbalance systems}, such as the
nucleonic superfluids under isospin asymmetry or strong magnetic
fields~\cite{2016PhRvC..93a5802S,2014PhRvC..90f5804S}.

In addition to the study of Cas A behaviour,
Ref.~\cite{2016EPJA...52...44S} studied the dependence of the cooling
curves on the variations of the mass of the compact star model. It
was found that the heavy stars that are close to the maximal mass
will cool to lower temperatures much faster; therefore some refitting
of the parameters will be required to adjust these models to the case
of Cas A.  It was also found that low mass stars
$M\sim 1.1\, M_{\odot}$ remain warm over longer time scales and are
thus hotter than their heavy counterparts.

Some issues remain to be studied, for example the role of strange
quarks in the quark phases. These may influence the cooling in case
they are not involved in the pairing pattern. However, at
asymptotically high density, when the strange quark mass vanishes, the
color-flavor-locked (CFL) phase in which the pairing is ``perfect'',
in the sense that all colors and flavors are involved in the
condensate, is the ground state of matter~\cite{1999NuPhB.537..443A}.
Because the CFL phase has very low heat capacity as well as neutrino
emissivity strong modifications are expected only in the case where
imperfect CFL phases arise, such as the gapless CFL
phase~\cite{2008RvMP...80.1455A}.  In any event, the low number of
strange quarks can render the process on strange quarks subdominant.


\section{Transport phenomena}
\label{sec:transport}

The treatment of transport phenomena in dense QCD phases of neutron
stars requires the knowledge of excitation spectra of quarks near
their Fermi-surface. The pairing patterns in the color and flavor
spaces are essential in determining the content of excitations,
especially those that are ungapped.  At large densities the dominant
CFL phase is a superfluid, whose excitation spectrum contains only
phonons in analogy liquid $^4$He, see
Refs.~\cite{2002PhRvC..66a5802S,2007JCAP...08..001M,2005JHEP...09..076M}.
As the density is lowered to the densities relevant to compact stars
the precise from of pairing becomes uncertain. The most robust phase
at this density is the the two-flavor color-superconducting (2SC)
phase, which pairs up and down quarks in a color antitriplet
state. The blue color is either unpaired or undergoes single-flavor
and color pairing with very small gaps of the order of
keV~\cite{1984PhR...107..325B}. In addition to the bulk fermions, the
2SC phase can contain color-magnetic flux tubes which will carry
partially the magnetic field through the color
superconductor~\cite{2010JPhG...37g5202A}. The flux tubes can
effectively scatter bulk fermions and, therefore, contribute to the
transport and relaxation in a certain parameter range, which we will
discuss below.

\subsection{Bulk fermions}

The bulk transport coefficients of the 2SC phase were computed in
Ref.~\cite{2014PhRvC..90e5205A} using the Boltzmann equation in the
transport relaxation approximation.  The matrix elements of the
scattering processes were computed to the leading order in the
electromagnetic and strong coupling ($\alpha$ and $\alpha_s$).  The
screening of these interactions were taken in the Hard-Thermal-Loop
approximation. In addition it was assumed that the matter is strongly
degenerate with temperature $T\ll \mu_q$, where $\mu_q$ is the
chemical potential of quarks. An important feature of the interactions
is the mixing of the gauge fields into a linear combination of
electromagnetism and gluonic fields, called $\tilde Q$ and $X$. The
first field is unscreened in the 2SC phase and penetrates unimpeded in
the bulk; the second field is Meissner screened over the distances of
the order of the penetration depth.

The ungapped fermions of the 2SC phase are the blue up ($u$) and down
($d$) quarks and electrons.  Ref.~\cite{2014PhRvC..90e5205A} found the
following key features of the transport in the 2SC phase: the thermal
conductivity and shear viscosity are dominated by the blue-$d$ quarks
for sufficiently low temperatures. At higher temperatures the
electrons take over the dominant role. This transition for thermal
conductivity occurs at $T/\mu_q\sim \alpha/7.7\sim 10^{-3}$, so most
of the temperature range of interest for neutron stars is in the
blue-$d$ dominated regime with the scaling
$\kappa_{bd}= 0.00617/(T/\mu_q)$.  For the shear viscosity the
crossover from blue-$d$ to electron domination occurs at
$T/\mu_q\sim 10^{-5}$, so electrons become dominant at relatively low
temperature $T\sim 10$\,keV; the shear viscosity scales as
$\eta_e\simeq 0.00231 (\mu_q^4/5\pi^2) \alpha^{-5/3}T^{-1}
(T/\mu_q)^{-2}$.
On the other hand the electrical conductivity of the rotated
$\tilde{Q}$ charge is completely dominated by the electron, because
the blue-$d$ quarks are $\tilde Q$ neutral, whereas blue-$u$ quarks
are much less abundant; the electrical conductivity of electrons is
given by
$\sigma_e = 0.0433 (\mu_q^2e^2/\pi^2)\alpha^{-5/3}T^{-1}
(T/\mu_q)^{-2/3}$;
see Ref.~\cite{2014PhRvC..90e5205A} for the expressions of less
dominant contributions. It was also shown that $SU(2)_{rg}$ gluons
do not contribute to the transport coefficients in 2SC quark matter
significantly.

\subsection{Color-magnetic flux-tubes}

The magnetic field will penetrate the 2SC phase in form of quantized
color-magnetic flux tubes because  for realistic values of parameters
the 2SC phase is a type-II $X$-charge
superconductor~\cite{2010JPhG...37g5202A}. The corresponding
Ginzburg-Landau parameter is given by 
\bea
\kappa_{\rm 2SC} \approx 11\frac{\Delta}{\mu_q} \ge \frac{1}{\sqrt{2}}.
\eea
where $\Delta $ is the gap in the red-green channel. 
The critical field for formation of flux tubes is very high,
$H_{c1} \sim 10^{17}$~G compared to typical field of a neutron stars,
but as the 2SC phase nucleates, domains of superconducting matter
will form, which will be Meissner screened. This will leave the normal
regions with intense magnetic fields above the average value. Once
these regions eventually undergo superconducting phase transition the
magnetic field intensity within them will be larger than $H_{c1}$ with
$X$ flux trapped in the form of flux tubes.  The detailes of the
configurations and density of the $X$-tubes is not known and depends,
among several things, on the dynamics of the phase
transition. However, their average density is expected to be of the
order of that of the ordinary flux tubes in the hadronic core by
simple flux conservation. 

\begin{figure}[h]
\centering 
\sidecaption 
\boxed{
\includegraphics[width=6cm]{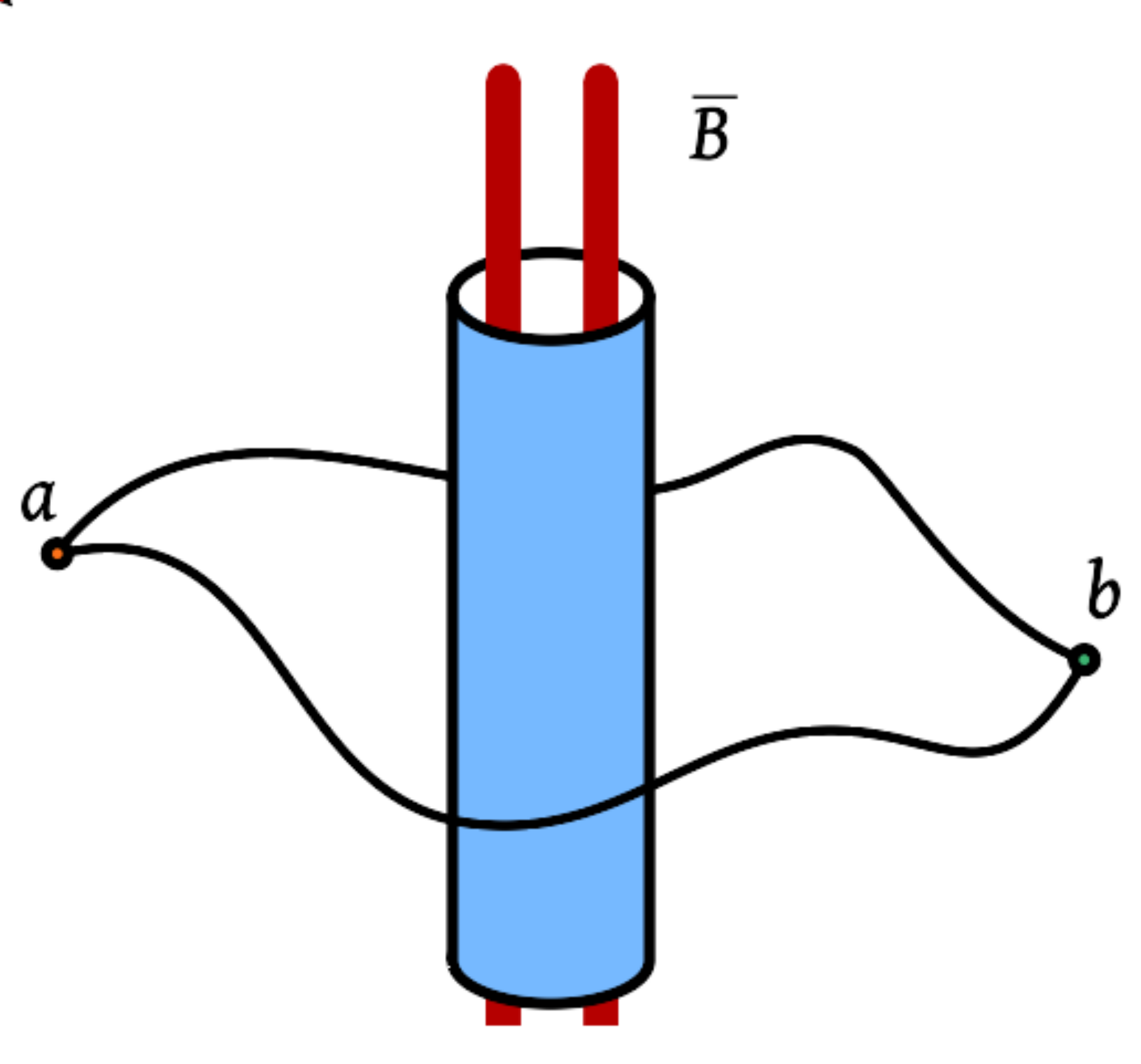}
}
\caption{Illustration of the Aharonov-Bohm effect: an electron or blue
  $u$-quark acquires a phase shift as it passes from point $a$ to
  point $b$ around a color-magnetic flux tube (shown as the cylindrical
  slab). }
\label{fig-3}       
\end{figure}
The Aharonov-Bohm scattering cross-section of gapless fermions with
the $X$ flux tubes was calculated in terms of the cross-section given
by 
\begin{equation}
 \frac{d\sigma}{d\vartheta} = 
\frac{\sin^2(\pi\tilde\beta)}{ 2\pi k\sin^2(\vartheta/2)},
\label{AB-scattering}
\end{equation}
where $k$ is the momentum in the plane perpendicular to the flux-tube,
$\vartheta$ is the scattering angle, $\tilde\beta= {q_p}/{q_c},$ $q_p$
is the charge of the scattering particle and $q_c$ is the charge of
the condensate whose winding by a phase of $2\pi$ characterizes the
flux-tube. The values of $\tilde \beta$ are given by
\bea 
\tilde\beta = {\rm diag}\Bigl(
   \frac{1}{2}+\frac{3}{2}\sin^2\varphi,  
   \frac{1}{2}-\frac{3}{2}\sin^2\varphi, 
   \frac{1}{2}-\frac{3}{2}\sin^2\varphi,
   \frac{1}{2}+\frac{3}{2}\sin^2\varphi, 
  -1+3\sin^2\varphi, 
   -1,  
   -3\sin^2\varphi \Bigr) 
\label{ABfactors}
\eea
in the basis $\psi = (ru,gd,rd,gu,bu,bd,e^-),$ where $rgb$ refer to
red, green, and blue colors, $e^-$ refers to electron, and 
the mixing angle is defined as $\cos\phi = \sqrt{3}g/\sqrt{e^2+3g^2}$,
where $e$ and $g$ are the electromagnetic and strong coupling constants.

The relaxation rate for particles of species $i$ scattering off flux
tubes of area density $n_v$ is then given by ~\cite{2010JPhG...37g5202A}
\bea 
\tau^{-1}_{if} = \frac{n_v}{p_{Fi}}  \sin^2(\pi\tilde\beta_i) \ . 
\label{tauinv-flux}
\eea 
where $n_v$ is the number density of color-magnetic flux tubes,
$p_{Fi}$ is the Fermi momentum of fermion $i$.  This result
corresponds to the standard form for relaxation rate known for
classical gases $\tau^{-1}=c n_v \sigma $, where $c=1$ is the speed of
the particles, $n_v$ is the density of scattering centers, and
$\sigma\propto \sin^2(\pi\tilde\beta)/p_F$ is the cross section of
Aharonov-Bohm scattering. Eq.~\eqref{tauinv-flux} sets the scale for
the thermal relaxation of the blue quarks and electrons in the 2SC
phase.  Note however that the blue-$d$ quark, has zero Aharonov-Bohm
interaction with the flux-tubes because $\tilde\beta=0$.  The other
two, the electron and blue-$u$ quark, have the same Aharanov-Bohm
factors, see Eq.~\eqref{ABfactors}, but different Fermi momenta.

Ref.~\cite{2010JPhG...37g5202A} argued that because the magnetic field
in hadronic envelope is below the lower critical field $H_{c1}$ for
$X$-flux-tubes into 2SC quark matter, the trapped flux-tubes will be
pulled out of the quark core. The force pulling the flux-tubes is
balanced by the mutual friction force on the moving flux-tube due to
its Aharonov-Bohm interaction with electrons and blue-$u$ quarks.
Apart from the rearrangement of the magnetic field, which is
characterized by time-scales comparable to pulsar life-times, the flux
expulsion has other physical significance: it could be a significant
source of heating of the lower base of the nuclear mantle, and
therefore the inner core (which is roughly isothermal after $10^3$
yr).  Internal heating is known to affect the thermal evolution of
compact star in the later stages of their thermal evolution (photo
cooling era).

Consider now the combined effect of scattering of a fermion off other
fermions and flux-tubes. If there are several scattering channels then
the net relaxation time for fermion species $i$ is given by the
\begin{equation}
\tau_i^{-1} =  \tau_{i,v}^{-1} + \sum_{j} \tau_{ij}^{-1}
\label{tau-sum}
\end{equation}
where $\tau^{-1}_{i,v}$ and $\tau^{-1}_{ij}$ are the fermion-flux-tubes
and fermion-fermion relaxation rates, respectively. In a first
approximation one can assume that the fermion-fermion and
fermion--flux-tubes scattering do not interfere. Then, it is clear that
the fermion--flux-tubes scattering shortens the mean-free-path of the
partciles and, therefore, suppresses the transport coefficients.

With the picture above, some conclusions can be drawn for the
individual transport coefficients~\cite{2014PhRvC..90e5205A}.  Because
the dominant contribution to the thermal conductivity is from blue-$d$
quarks, which have zero Aharanov-Bohm scattering with vortices, the
scattering off the flux-tubes has no significant effect on the thermal
conductivity. The electrical conductivity and shear viscosity are
dominated by electrons where the interactions are dominated by the
$\tilde{Q}$ interaction.

The electron-flux-tubes relaxation rates for electrical conductivity
and shear viscosity are approximately the same, as these involve the
momentum relaxation up to a constant of order $1$.  Clearly the
electron-flux-tubes sattering becomes important, when its relaxation
time is comparable to the fermion-fermion relaxation time.  According
to the estimates of Ref.~\cite{2014PhRvC..90e5205A} flux-tubes can lower
the magnitude of transport coefficients for realistic values of the
external magnetic field.

\section{Spin-up and spin-down induced phase transitions}
\label{sec:spin}

The studies of compact stars with quark cores in a wide range of
rotation periods from the Keplerian limit down to the static limit is
of great interest in the interpretation of the phenomena related to
the physics of pulsars. Compact stars that are born rapidly rotating
evolve adiabatically by losing energy to gravitational waves and
electromagnetic (to lowest order, dipole) radiation. Such evolution
can be approximated by a sequence of configurations with constant rest
mass and varying spin. The massive rapidly rotating configurations may
not have a stable static limit. Consequently, the spin-down evolution
can lead to the formation of a black hole when the support through
centrifugal force cannot balance the pull of gravity anymore.  The
configurations which do not have static counterparts are known as {\it
  supramassive configurations}~\cite{1994ApJ...422..227C}.  The
time-reversed counterpart of the spin-down scenario is also realized
in nature. According to the formation scenario of the millisecond
pulsars, these are spun-up to millisecond frequencies by accretion
from a companion. In this case the evolution involves some mass
transfer via accretion to the compact star, therefore the rest mass of
the star is not a constant anymore. However, the changes in the rest
mass during the accretion are not significant and do not change the
qualitative picture. Therefore, the changes in the structure of the
compact stars as the star spin increases with time can be considered, in
a first approximation, as ``time reversed'' evolution, which does not
require additional physical input, such as the mass accretion rate
from the companion.

How the spin evolution of hybrid stars, i.\,e., stars with
quark-matter cores surrounded by a nuclear envelope, may differ from
the evolution of purely nucleonic stars?  The most dramatic effect is
the spin-down or spin-up induced phase transition(s) between the
hadronic and quark matter phases. The reason is that rapidly rotating
stars are {\it less compressed} and are effectively more
``low-density'' objects than their static counterparts. Changes in the
spin of the star induce changes in its density profile, which in turn
can lead to density-induced phase transitions from quark to hadronic
matter and back.  The physics of spin-up and -down induced phase
transitions in {\it non-superconducting} quark matter was first
discussed discussed by Glendenning, Pei and Weber in
~Ref.~\cite{1997PhRvL..79.1603G} and subsequently elaborated in many
studies, see e.~g.,
Refs.~\cite{2001ApJ...559L.119G,2003NuPhA.715..831G,2006A&A...450..747Z,2009MNRAS.392...52A,2013arXiv1307.1103W}.

Rotating compact stars containing color superconducting cores were
considered initially in Ref.~\cite{2008PhRvD..77b3004I}.  The color
superconducting matter was taken to be in the crystalline phase. The
case where the quark matter pairs either in the 2SC or the CFL phases
was subsequently studied in Ref.~\cite{2013A&A...559A.118A} on the
basis of an equation of state constructed to reproduce the nuclear and
quark matter phenomenologies as well as the astrophysical constraint
on the value of the maximum mass of a compact star $M/M_{\odot}\ge 2$.
Below, our discussion follows the study of
Ref.~\cite{2013A&A...559A.118A}, where the equation of state of a
hybrid star was constructed assuming a sharp interface between the
phases. This is the case when the surface tension of the interface is
not too small. The hadronic matter description is based on
relativistic density functional of nuclear matter with
density-dependent couplings, see
Refs.~\cite{2005PhRvC..71b4312L,2013PhRvC..87e5806C} for the
parameterization and further details. The quark  matter equation of
state is based on the Nambu--Jona-Lasinio model which includes in
addition to common interactions also repulsive vector interactions as
given in Ref.~\cite{2012A&A...539A..16B}.
\begin{figure}[t]
\centering 
\sidecaption 
\includegraphics[width=7cm]{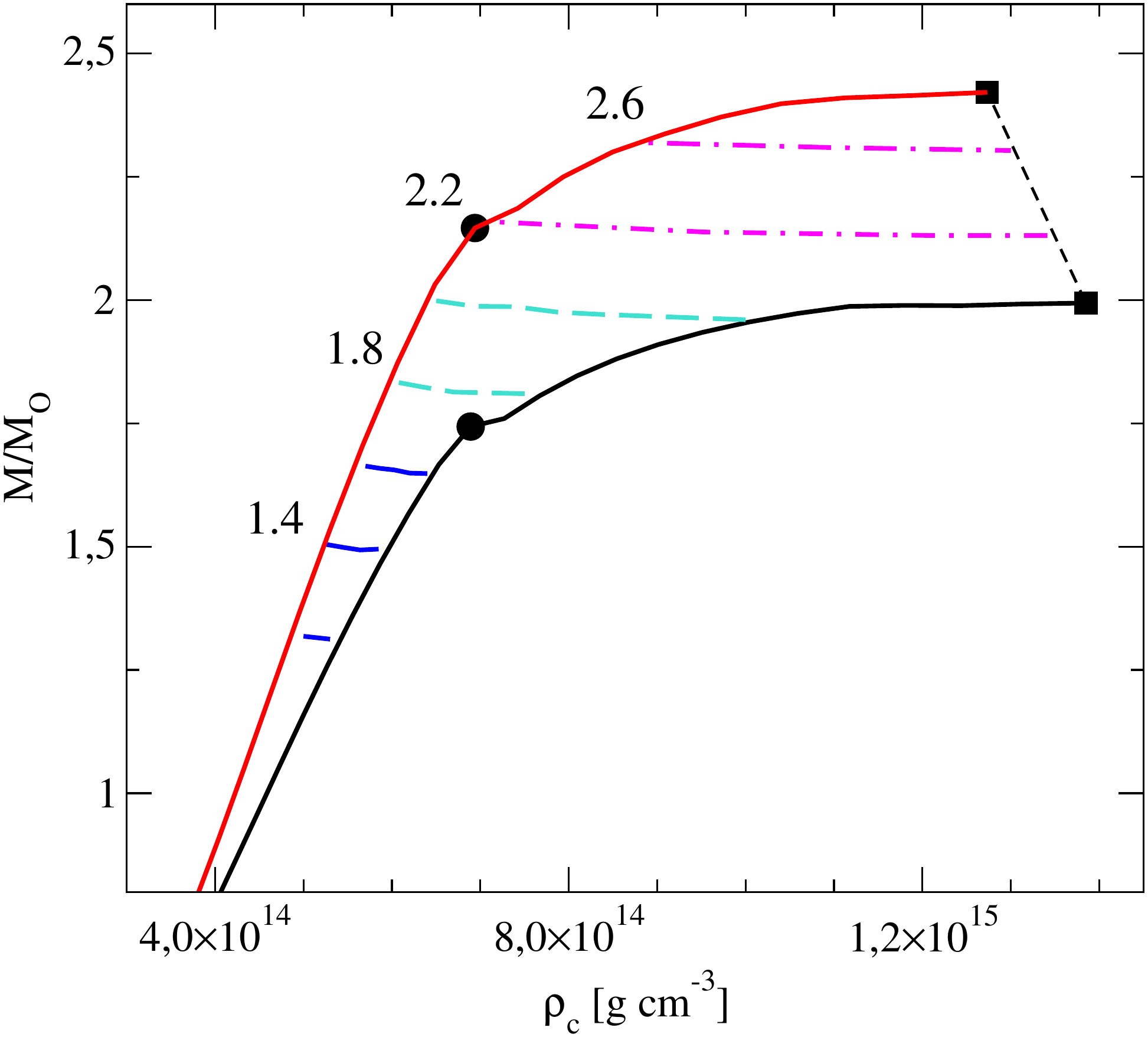}
\caption{ Dependence of masses of hybrid compact stars on the central
  density. The lower and upper solid lines corresponds to non-rotating
  and maximally rotating stars.  The dashed and dash-dotted lines show
  stars with a fixed rest masses, as labeled in the plot, with the
  mass step between two sequential lines equal 0.2. Among these the
  dashed-dotted lines correspond to the class of the stars which do
  not have static counterparts. The squares show the maximal mass
  stars, which are connected by the line at which the stability region
  terminates.  The stars located to the right of dots are hybrid stars
  that contain quark matter. All mass units are given in units of the
  solar mass $M_{\odot}$.  }
\label{fig-4}       
\end{figure}
The mass vs central density relations for hybrid stars are shown in 
Fig.~\ref{fig-4} in the static and Keplerian limits. These limits can
be connected by constant {\it rest mass} lines (near horizontal lines 
in Fig.~\ref{fig-4}), which correspond to the scenarios of spin-up or
spin-down of the stars without mass transfer. There are three classes of
constant rest mass sequences in the case of hybrid stars: 
\begin{enumerate}
\item The well-known {\it supramassive} compact stars which are stable
  only due to the (uniform) rotation (dashed-dotted lines in
  Fig.~\ref{fig-4}) . They collapse to a black hole as they spin down
  below certain critical angular velocity; the instability sets-in in the
  region to the right of the line connecting the maximal masses of
  non-rotating and maximally fast rotating compact stars in
  Fig.~\ref{fig-4}.
\item The second class can be defined as {\it transitional} compact
  stars (dashed lines in Fig.~\ref{fig-4}). These are purely nucleonic
  in the limit of maximally fast rotation and are hybrid in the static limit.
  This implies that there is a phase transition from nucleonic phase
  to the quark matter phase {\it at the center} of the star as the star
  spins  down or up. Note that phase transition take  also place within the 
  supramassive sequences which contain quark matter already in the limit
  of maximal rotation.  Because the central density and the density
  profiles of these stars change with their spin there is a continuous
  conversion of quark matter to nucleonic matter and back depending on
  whether the star spins down or up.
\item The third class of stars do not contain quark matter in either
  limit, i.e., these are purely nucleonic as the central densities are
  not sufficiently large for quark matter to form (solid lines in
  Fig.~\ref{fig-4}). The masses of these stars are below
  $1.7M_{\odot}$ for the equation of state used in our example.
\end{enumerate}
\begin{figure}[t]
\centering 
\sidecaption 
\includegraphics[width=7cm,clip]{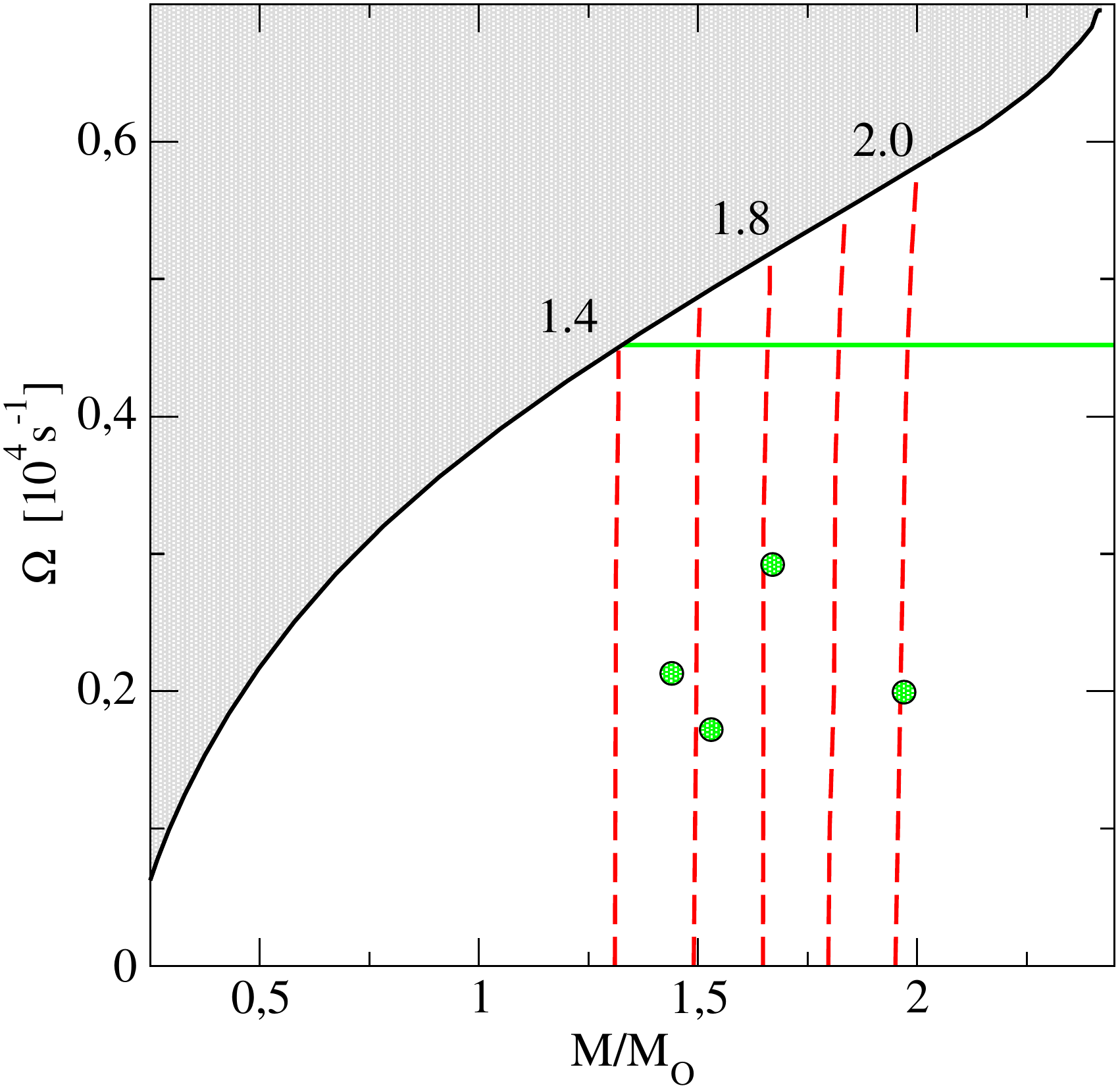}
\caption{ Frequency vs mass diagram for compact stars. Only the most
  massive object J1614-2230 is massive enough to contain quark
  matter. The solid line indicates the Keplerian frequency as a
  function of mass; the area above the line is excluded for uniformly
  rotating stars.
}
\label{fig-5}       
\end{figure}
Figure \ref{fig-5} displays the frequency vs mass diagram for
compact stars, with dots corresponding to objects for which the mass
and the spin frequency are measured, see
Ref.~\cite{2013A&A...559A.118A} for details. The shaded region is
inaccessible to our models in the case of uniform rotation. The 
horizontal line shows the spin frequency of the fastest rotating 
\begin{figure}[hbt]
\centering 
\begin{minipage}{6.0cm}
\includegraphics[width=6cm,clip]{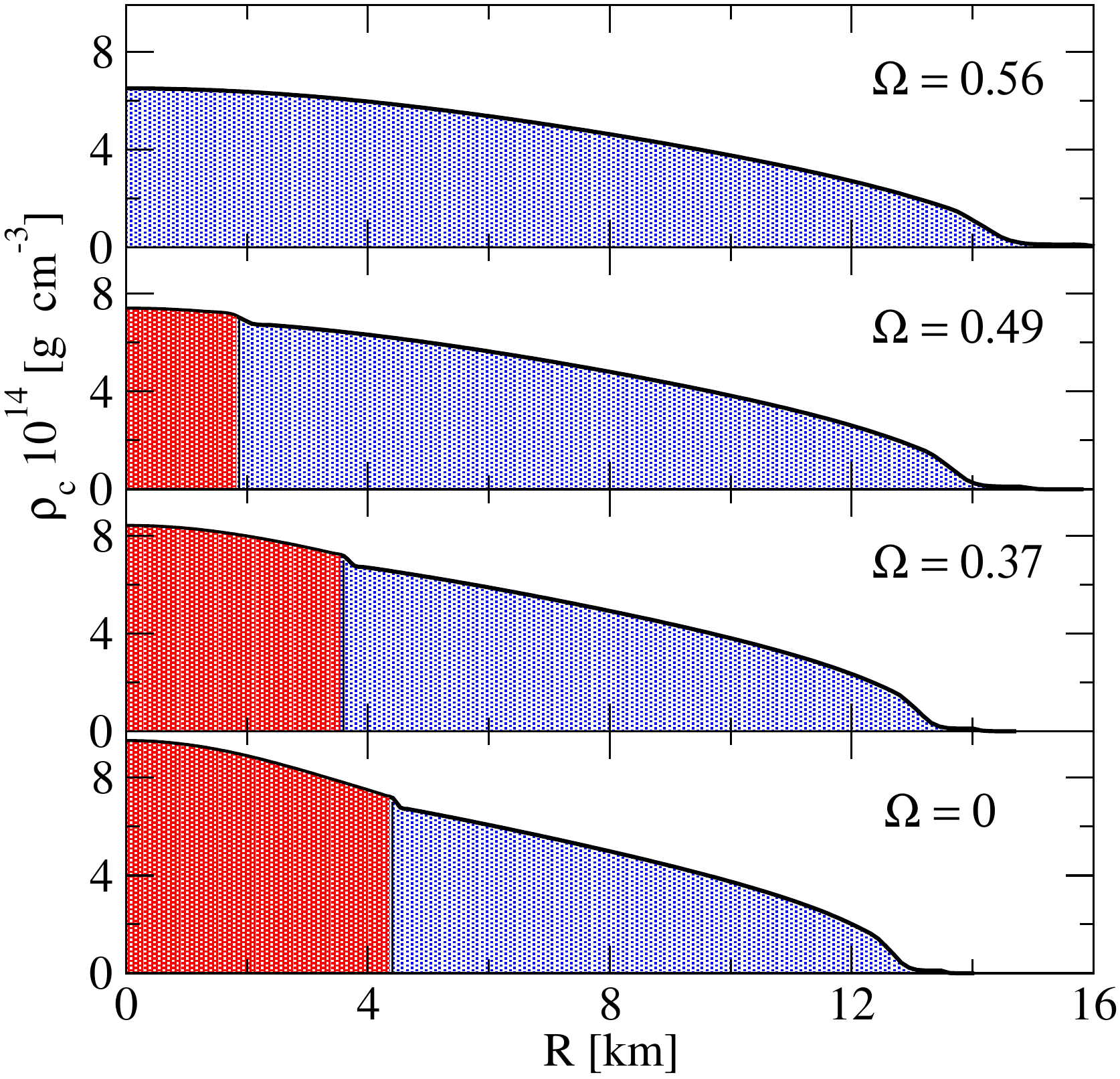}
\end{minipage}
\begin{minipage}{6.0cm}
\includegraphics[width=6cm,clip]{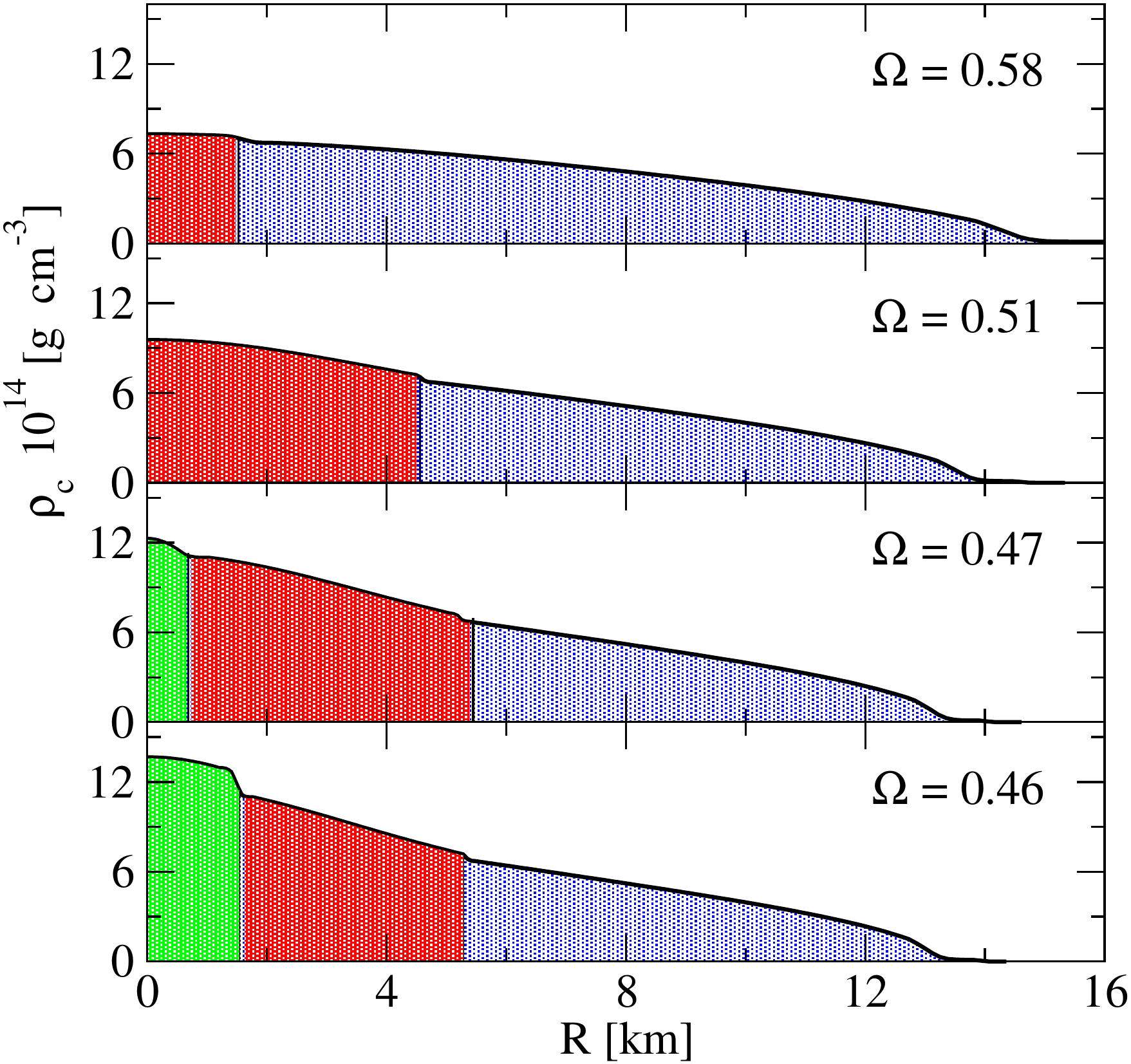}
\end{minipage}
\caption{Equatorial density profiles of a rotating compact stars with 
  rest masses 2.2 $M_{\odot}$ (left panel) and 2.4 $M_{\odot}$ (right 
  panel). The low-density region corresponds to nucleonic matter 
  (blue), which is followed by the 2SC phase (red) and CFL phase 
  (green).  The frequencies are given in units of $10^4$ s$^{-1}$. }
\label{fig-6}       
\end{figure}
pulsar J1748-2446, for which the mass is unknown. For present equation
of state this object's mass needs to be above $1.4M_{\odot}$, which
illustrates the possibility of constraining the equations of state of
dense matter by future measurements of the spin of sub-millisecond
pulsars.

Figure \ref{fig-6} shows the dependence of the equatorial density 
profile of 2.2 $M_{\odot}$ (left panel) and 2.4 $M_{\odot}$ (right panel)
rest-mass stars on the internal radius for four spin frequencies in
the range $0\le \Omega\le 5.6 \times 10^3$ Hz.  The upper panels in
both cases show the stars rotating at the Keplerian limit.  As
expected, the region occupied by the 2SC phase becomes larger as the
star's spin is reduced.  Eventually, the CFL phase may form at the
center of the most massive stars.  Because the phase transition from
nucleonic phase to the 2SC phase and from the 2SC to the CFL phase is
first order, both are accompanied by jumps in the density, as seen in
Fig. \ref{fig-6}

\section{Final remarks}
\label{sec:conclusions}

The exploration of the dense QCD via the astrophysics of compact stars
is an actively pursued subject. We expect substantial observational
progress in the studies of compact stars in the upcoming years,
including further measurements of the masses of pulsars, the radii of
neutron stars, and eventually the gravitational wave signals from the
mergers of two neutron stars or neutron stars and black holes. This
motivates strongly further studies of dense matter in all its facets,
including the physics that goes beyond the equation of state. Better
understanding of the transport coefficients of dense QCD, weak
interactions in dense matter, and dynamics of phase transitions would
largely improve our modelling of astrophysical phenomena that are
relevant for the current observational programs.

\section*{Acknowledgments}

I would like to thank my collaborators M.~Alford, L.~Bonanno,
G.~Colucci, X.-G.~Huang, H.~Nishimura, and D.~Rischke, for their
contribution and insight into the research described above.  This work
was supported by the Deutsche Forschungsgemeinschaft (Grant No. SE
1836/3-2), by the Helmholtz International Center for FAIR, and by the
NewCompStar COST Action MP1304.


\end{document}